\documentclass{article}
\usepackage{amsfonts}
\usepackage{graphicx}

\evensidemargin=\oddsidemargin
\def\Title#1#2#3{%
    \baselineskip=18pt
    \begin{center}
          {\large\bf{#1} \\ }
          \bigskip\bigskip
          {#2} \\
          {#3} \\
    \end{center}}
\long\def\Abstract#1{%
         \bigskip
         \parbox{0.93\textwidth}{%
                 \begin{center}
                       {\bf Abstract} \\
                 \end{center}
                 \medskip{\baselineskip=14pt #1}
                 \vss}
         \bigskip}

\makeatletter
\renewcommand{\section}%
 {\@startsection{section}{1}{0pt}%
  {-3.25ex plus -1ex minus -.2ex}{1.5ex plus .2ex}%
  {\vspace*{5mm}\raggedright\large\bf }}
\renewcommand{\subsection}%
 {\@startsection{subsection}{2}{0pt}%
  {-2.25ex plus -.5ex minus -.2ex}{-1.5ex plus -.2ex}{\bf }}
\renewcommand{\subsubsection}%
 {\@startsection{subsubsection}{3}{0pt}%
  {-1.25ex plus -.2ex minus -.1ex}{-1.2ex plus -.2ex}{\bf }}

\makeatother

\begin{document}

\Title{Some features of the extended phase space approach\\
 to quantization of gravity}
{T. P. Shestakova\footnote{E-mail: {\tt shestakova@sfedu.ru}}}%
{Department of Theoretical and Computational Physics,
Southern Federal University,\\
Sorge St. 5, Rostov-on-Don 344090, Russia}

\Abstract{In this paper, I emphasize those features of the extended phase space approach to quantization of gravity that distinguish it among other approaches. First of all, it is the conjecture about non-trivial topology of the Universe which was supported by Wheeler, Hawking and other founders of quantum gravity. However, this conjecture appears to be in contradiction with the assumption about asymptotic states that is used in the path integral quantization of gauge theories. The presence of asymptotic states ensures gauge invariance of the theory, but, in the case of gravity, the states exist only in asymptotically flat spacetimes, that limits possible topologies. Then we have two ways. The first way is to consider only asymptotically flat spacetimes. In fact, it reduces quantum gravity to quantum field theory on a given background. The second way is to reject the assumption about asymptotic states. In the case of non-trivial topology, one cannot cover the whole spacetime with the only coordinate system. One has to introduce various reference frames fixed by different gauge conditions in different spacetime regions. The Hamiltonian describing a gravitating system will depend on gauge conditions. It leads to the conclusion that unitary evolution may be broken down. This conclusion cannot be obtained in approaches based on the Wheeler -- DeWitt equation or making use of the assumption about asymptotic states. The assessment of this conclusion is given.}

\section{Introduction}
The founders of quantum geometrodynamics often spoke that the Universe may have a non-trivial topology. So, in 1955, yet before quantum geometrodynamics was formulated in the seminal paper of DeWitt \cite{DeWitt}, John Wheeler had put forward an idea of fluctuations of spacetime geometry \cite{Wheeler}, that later became known as spacetime foam.

In the volume published in 1979 and devoted to an Einstein centenary, Hawking wrote that one would expect that quantum gravity would allow all possible topologies of spacetime, and it seems that taking into account various topologies may give the most interesting effects \cite{Hawking}.

However, the conjecture about non-trivial topology of the Universe appears to be in contradiction with the assumption about asymptotic states that is used in the path integral quantization of gauge theories.

As it is well-known, there are two basic approaches to quantization: the canonical approach relying on Hamiltonian formalism, and the path integral approach. In the canonical approach, spacetime topology is restricted by the product of the real line with some three-dimensional manifold,
${\mathbb R}\times\Sigma$. In quantum field theory, the path integral approach was originally used for construction of $S$-matrix, that implies that particles in initial and final (asymptotic) states are outside the interaction region. In its turn, it means that the path integral is considered under asymptotic boundary conditions which exclude non-physical degrees of freedom in initial and final states. The asymptotic boundary conditions ensure gauge invariance of the path integral and, therefore, gauge invariance of the whole theory. However, in the case of gravity, the assumption about asymptotic states is true only in asymptotically flat spacetimes. Let us note, also, that the main goal is not to construct $S$-matrix, but to quantize the full gravitational theory.

We come to the conclusion that the both canonical and path integral approaches do not admit an arbitrary spacetime topology. If one restricts topology to the product $\mathbb R\times\Sigma$ or to asymptotically flat spacetimes, one would get a quantum field theory on some fixed background, but not a full quantum gravity.

What would be if one refuses the assumption about asymptotic states? In this case one cannot prove gauge invariance of the theory. The Wheeler -- DeWitt equation, which is believed to express this gauge invariance, would lose its sense. But one can derive from the path integral a Schr\"odinger equation for a wave function of the Universe instead. The Schr\"odinger equation is expected to maintain its fundamental sense \cite{Shest1}.

The Wheeler -- DeWitt equation is a direct consequence of the Dirac quantization scheme for constrained theories. In fact, most of approaches to quantization of gravity have been elaborated to be consistent with the Dirac quantization scheme. Dirac was excited by the role that Hamiltonian formalism had played when quantum mechanics had been created. He wrote in his ``Lectures on quantum mechanics'' \cite{Dirac1} that
\begin{quote}
``\ldots if we can put the classical theory into the Hamiltonian form, then we can always apply certain standard rules so as to get a first approximation to a quantum theory.''
\end{quote}

Nevertheless, the construction of the Hamiltonian formalism for constrained systems was not a trivial task. It is notorious that, for these systems, one cannot express all generalized velocities in terms of momenta to wrote a Hamilton function by the usual rule
\begin{equation}
\label{Ham_func}
H=p_a\dot q^a+\pi_{\alpha}\dot\lambda^{\alpha}-L.
\end{equation}
Here, all degrees of freedom of the theory are naturally divided into two groups: the so-called ``physical'' variables $\{q^a\}$ and their conjugate momenta $\{p_a\}$, and ``non-physical'' (or gauge) degrees of freedom $\{\lambda^{\alpha}\}$ and their momenta $\{\pi_{\alpha}\}$. Equations for the latter ones
\begin{equation}
\label{mom}
\pi_{\alpha}=\frac{\partial L}{\partial\dot\lambda^{\alpha}}=0
\end{equation}
do not enable us to express the velocities $\dot\lambda^{\alpha}$ in terms of momenta.

Dirac is believed to find the solution to the problem by introducing the following two postulates:
\begin{itemize}
\item One should add a linear combination of constraints $\{\varphi_{\alpha}\}$ to the Hamiltonian:
\begin{equation}
\label{Dir_Ham}
H=H_0+\lambda^{\alpha}\varphi_{\alpha}.
\end{equation}
\item When quantizing, the constraints in the operator form become conditions imposed on the state vector:
\begin{equation}
\label{quant_constr}
\varphi_{\alpha}|\Psi\rangle=0.
\end{equation}
\end{itemize}

Why the rules (\ref{Dir_Ham}) and (\ref{quant_constr}) are postulates? They cannot be derived from other fundamental physical statements, cannot be justified by the reference to the correspondence principle, etc. Moreover, these postulates have never been verified by any physical experiments, while very successful theories, confirmed experimentally, are based on different methods. For example, quantum electrodynamics is based on Lagrangian formalism and perturbation theory. Ironically, the Dirac approach is used only in various attempts to quantize gravity, in other words, in the sphere where, until now, we have not got any experimental data.

Meanwhile, the development of quantization methods gave a hint how Hamiltonian dynamics can be constructed differently. In the path integral quantization of gauge theories, the gauge invariant action of an original theory is replaced by an effective action which includes gauge fixing and ghost terms. A gauge condition can be chosen in such a way that it would introduce missing velocities into the effective Lagrangian. An example is given by the Lorentz gauge in electrodynamics,
\begin{equation}
\label{eff_act}
S_{ED}\rightarrow S_{eff}=\int d^4x\left({\cal L}_{ED}+{\cal L}_{gf}+{\cal L}_{ghost}\right);
\end{equation}
\begin{equation}
\label{Lor_gauge}
{\cal L}_{gf}=\pi\partial_{\mu}A^{\mu}=\pi\left(\dot A^0+\partial_iA^i\right).
\end{equation}
Here, $\pi$ is a Lagrange multiplier, that is, at the same time, a momentum conjugate to $A^0$. It is easy to see that, if one substitute the effective Lagrangian into (\ref{Ham_func}), the terms with $\dot A^0$ vanish, and the Hamilton function can be constructed according the usual rule (\ref{Ham_func}).

In Section 2, the main features of the new formulation of Hamiltonian dynamics, which is an alternative to the generalized Hamiltonian dynamics of Dirac, are discussed. In Section 3, we shall turn to quantization and consider a Schr\"odinger equation derived from the path integral without asymptotic boundary conditions. It is worth attention that the Hamilton operator in the Schr\"odinger equation corresponds (up to operator ordering) to the Hamilton function in extended phase space. Since no asymptotic boundary conditions were imposed, the Schr\"odinger equation appears to be gauge dependent. This very circumstance enables us to speculate how non-trivial spacetime topology can be taken into account in this approach, and where it is going. Indeed, in the case of non-trivial topology one cannot introduce only one reference frame in the whole spacetime, but has to introduce various reference frames in different spacetime regions. The formalism where the Schr\"odinger equation depends on chosen gauge conditions (a reference frame) seems to be suitable to tackle the problem. This is discussed in Section 4, while some conclusions are drown in Section 5.

\section{Hamiltonian dynamics in extended phase space}
Since we deal with the effective action, we should work in extended phase space, that includes, on the equal footing, physical, gauge and ghost degrees of freedom. It is natural to refer to this formulation of Hamiltonian dynamics as {\it Hamiltonian dynamics in extended phase space} and, since it is a prerequisite of quantization, the proposed approach has been called {\it the extended phase space approach to quantization of gravity}.

Eq. (\ref{Ham_func}) should be correctly rewritten as
\begin{equation}
\label{Ham_gfunc}
H=p_a\dot q^a+\pi_{\alpha}\dot\lambda^{\alpha}
 +\bar{\cal P}_{\alpha}\dot\theta^{\alpha}+\dot{\bar\theta}_{\alpha}{\cal P}^{\alpha}-L.
\end{equation}
Here $\{\bar\theta_{\alpha},\theta^{\alpha}\}$ are pairs of Faddeev -- Popov ghosts,
$\{{\cal P}^{\alpha},\bar{\cal P}_{\alpha}\}$ are their conjugate momenta and we use the ordering rule that $\bar{\cal P}_{\alpha},\bar\theta_{\alpha}$ are written on the left, while
${\cal P}^{\alpha},\theta^{\alpha}$ are written on the right. Correspondingly, when obtaining the Hamilton equations, we take left derivatives of $\bar{\cal P}_{\alpha},\bar\theta_{\alpha}$ and right derivatives of ${\cal P}^{\alpha},\theta^{\alpha}$. This enables one to avoid needless multipliers $(-1)$ as a result of commuting Grassmannian variables.

Using the effective action means that the variation procedure gives modified Einstein equations that includes additional terms resulting from the gauge fixing and ghost parts of the action. One should add gauge conditions and ghost equations to the modified Einstein equations, so that one comes to {\it the extended set of Lagrangian equations}.

The Hamiltonian set of equations in extended phase space is completely equivalent to the extended set of Lagrangian equations. The equivalence has been verified for models with a finite number of degrees of freedom (see, for example, \cite{Shest2}) as well as for the spherically symmetric gravitational model \cite{Shest3}), which has an infinite number of degrees of freedom. The proof of the equivalence is straightforward though it requires cumbersome calculations for some models. The equivalence implies that constraints, gauge conditions and ghost equations are Hamilton equations. Thus, the description of the dynamics appears to be as close as possible to the description of a system without constraints, while the constraints are preserved. They are modified just like other Einstein equations.

In the Dirac approach, the status of gauge variables is not clear. At first, he included them into phase space and into the definition of the Poisson brackets to obtain secondary constraints. But then, he wrote that these variables are not of interest, and, therefore, one could drop them out of the theory \cite{Dirac1}. After that, most researches consider them as redundant. Even in the approach of Batalin, Fradkin and Vilkovisky (BFV) \cite{BFV1,BFV2,BFV3}, who included gauge and ghost degrees of freedom into the definition of the path integral and introduced the very notion of extended phase space, their role is just auxiliary. Meanwhile, there exist a problem related with the choice of gravitational variables.

In the theory of gravity, different parameterizations of variables are used. The gravitational field can be represented by components of metric tensor or by the Arnowitt -- Deser -- Misner variables. From the viewpoint of the Lagrangian formalism, it is just a variable change,
\begin{equation}
\label{ADM}
g_{00}=\gamma_{ij}N^iN^j-N^2;\quad
g_{0i}=\gamma_{ij}N^j;\quad
g_{ij}=\gamma_{ij}.
\end{equation}
In theories without constraints, any variable change in the Lagrangian formalism corresponds to a canonical transformation in the Hamiltonian formalism. However, one can check that, in the Dirac approach, the change of variables (\ref{ADM}), which touches upon gauge variables, is not canonical, even if one includes gauge degrees of freedom into the definition of the Poisson brackets. The Poisson brackets of the lapse function $N$ and momenta $\Pi^{ij}$ conjugate to space components of the metric tensor $\gamma_{ij}$ is not zero \cite{KirKuz,Shest2}:
\begin{equation}
\label{Kir_Kuz}
\left.\{N,\Pi^{ij}\}\right|_{g_{\mu\nu},p^{\lambda\rho}}\ne 0.
\end{equation}
The change of variables (\ref{ADM}), which is absolutely legal in the Lagrangian formalism, leads to a contradiction from the viewpoint of the Dirac approach. At least, it means that Hamiltonian dynamics of Dirac is not completely equivalent to the original (Lagrangian) formulation of the Einstein theory.

As was shown in \cite{Shest2,Shest4}, the problem has been solved in the Hamiltonian formulation in extended phase space. Thanks to introducing the gauge fixing term into the effective action, the momenta $\Pi^{ij}$ are modified, that results in correct values of the Poisson brackets. It has been proved for the full gravitational theory that changes of variables like (\ref{ADM}) are canonical transformations in extended phase space.

Another problem, that has been also solved in the proposed approach to Hamiltonian dynamics, is construction of BRST generator. The effective action is not gauge invariant, but there exists a residual global invariance of the action revealed by Becchi, Rouet, Stora and Tyutin (BRST) \cite{BRS,Tyut}. In the Lagrangian formalism, BRST transformations for variables of the original theory coincide with gauge transformations. In the Dirac approach, gauge transformations are generated by constraints, but constraints do not generate correct transformations for gauge variables. (By ``correct'' transformations, I mean those that correspond to gauge ones in the Lagrangian formalism.) Dirac did not worried about gauge variables, we remember that he considered them as being not of interest. Batalin, Fradkin and Vilkovisky proposed a method of constructing BRST generator based on the constraints algebra \cite{BFV2,Hennaux}. It is not surprising that, as well as the constraints, this generator does not produce correct transformations for gauge variables.

Actually, the group of transformations generated by constraints is not the same as the gauge group in the Lagrangian formulation of general relativity, even more, the later group is closed while the former is open. Remarkably, but the creators of the BFV approach were aware of this circumstance. In \cite{BFV1}, Fradkin and Vilkovisky wrote:
\begin{quote}
``\ldots in this case [of the gravitational theory] the gauge transformations cannot be presented as canonical transformations in Hamiltonian theory\ldots and thus they differ from transformations generated by constraints.''
\end{quote}
Since the algebra of transformation generated by the constraints is open, an additional term appears in the BFV effective action for gravity, that corresponds to the four-ghosts interaction and would never arise in the Lagrangian formalism. Nevertheless, nobody paid a serious attention to this fact, since it concerned a non-physical (redundant) sector of the theory.

However, the global BRST invariance enables us to construct the BRST generator in accordance with the Noether theorem. In this case the generator gives correct transformations for all degrees of freedom including gauge ones. It has been demonstrated in \cite{Shest2,Shest3} for a model with a finite number of degrees of freedom and the spherically symmetric gravitational model. In the Noether theorem, the Lagrangian formalism is used, so that everything is in agreement.

In conclusion of this Section, let us write down the effective action for a model with a finite number of degrees of freedom:
\begin{equation}
\label{S_fin}
S=\int dt\left[\frac12 g_{ab}(N,q)\dot q^a\dot q^b-U(N,q)
 +\pi\left(\dot N-\frac{\partial f}{\partial q^a}\dot q^a\right)+N\dot{\bar\theta}\dot\theta\right].
\end{equation}
Here $\{q^a\}$ stands for physical degrees of freedom (as in (\ref{Ham_gfunc})), $g_{ab}$ is the metric of configurational space, which depends on a gauge variable $N$ (be it the lapse function or not), $U(N,q)$ is some potential, and a differential form of the gauge condition $N=f(q)$ is used. The Hamilton function in extended phase space is:
\begin{eqnarray}
\label{Ham_fin}
H&=&\frac12 g^{ab}p_ap_b+\pi p_a\frac{\partial f}{\partial q_a}
 +\frac12\pi^2\frac{\partial f}{\partial q^a}\frac{\partial f}{\partial q_a}
 -U(N,q)+\frac1N{\cal\bar P}{\cal P}\nonumber\\
&=&\frac12 G^{\alpha\beta}P_{\alpha}P_{\beta}+U(N,q)+\frac1N{\cal\bar P}{\cal P};
\end{eqnarray}
where
\begin{eqnarray}
\label{matrG}
G=\left(\begin{array}{cc}
\displaystyle\frac{\partial f}{\partial q^a}\frac{\partial f}{\partial q_a}
 &\frac{\partial f}{\partial q_a}\\
\displaystyle\frac{\partial f}{\partial q_a}& g^{ab}
\end{array}\right),
\end{eqnarray}
$Q^{\alpha}=\{N,q^a\}$; $P_{\alpha}=\{\pi,p_a\}$. In the next Section, we shall compare a Hamilton operator in the Schr\"odinger equation with the Hamilton function (\ref{Ham_fin}).

\section{The Schr\"odinger equation}
Let us quote Dirac again \cite{Dirac2}:
\begin{quote}
``Any dynamical theory must first be put in the Hamiltonian form before one can quantize it.''
\end{quote}
In the previous Section, we have outlined the new Hamiltonian formulation of the gravitational theory, so we can follow Dirac at that point and go to quantization.

Generalizing the Feynman method \cite{Feynman} of derivation of the Schr\"odinger equation for constrained systems, we come to the following equation:
\begin{equation}
\label{Schr_math}
i\frac{\partial\Psi(N,q,\theta,\bar\theta;t)}{\partial t}=H\Psi(N,q,\theta,\bar\theta;t).
\end{equation}
The Hamiltonian operator in this equation looks like
\begin{equation}
\label{Ham_op}
H=-\frac1{2M}\frac{\partial}{\partial Q^\alpha}
 \left(MG^{\alpha\beta}\frac{\partial}{\partial Q^{\beta}}\right)+U(N,q)+V[f]
 -\frac1N\frac{\partial}{\partial\theta}\frac{\partial}{\partial\bar\theta},
\end{equation}
where $M$ is the measure in the path integral, $V[f]$ is a quantum correction which is proportional to $\hbar^2$ and curvature of configurational space \cite{Cheng}. It is important that the equation (\ref{Schr_math}) is a direct mathematical consequence of the path integral with the effective action (\ref{S_fin}) without asymptotic boundary conditions\footnote{My student Roger I. Ayala O\~na suggested to refer to it as {\it the mathematical Schr\"odinger equation}, in contrast to the physical Schr\"odinger equation (\ref{Schr_phys}).}.

We can see that the Hamilton operator (\ref{Ham_op}) indeed corresponds (up to operator ordering) to the Hamilton function in extended phase space (\ref{Ham_fin}).

The wave function---a solution to Eq.(\ref{Schr_math})---is defined on the extended configurational space. The general solution to Eq.(\ref{Schr_math}) has the form
\begin{equation}
\label{gen_sol}
\Psi(N,q,\theta,\bar\theta;t)=\int\Psi_k(q,t)\delta(N-f(q)-k)(\bar\theta+i\theta)dk.
\end{equation}
The $\delta$-function fixes the gauge condition (up to a constant $k$). The function $\Psi_k(q,t)$ which depends only on physical variables $\{q^a\}$ contains information about a physical system. Substituting (\ref{gen_sol}) into (\ref{Schr_math}), we come to {\it the physical Schr\"odinger equation}:
\begin{equation}
\label{Schr_phys}
i\frac{\partial\Psi_k(q,t)}{\partial t}=H_{(phys)}[f]\Psi_k(q,t),
\end{equation}
\begin{equation}
\label{Ham_phys}
H_{(phys)}[f]=\left[-\frac1{2M}\frac{\partial}{\partial q^a}
 \left.\left(Mg^{ab}\frac{\partial}{\partial q^b}\right)+U(N,q)+V[f]\right]\right|_{N=f(q)+k},
\end{equation}
The wave function satisfying this equation describes geometry of the Universe from the point of view of an observer in the reference frame fixed by the given gauge condition.

The general solution (\ref{gen_sol}) is a result of decomposition by basis consisting of eigenfunctions of the operator $(N-f(q))$, which are $\psi_k=\delta(N-f(q)-k)$ in the coordinate representation. Indeed, one can write down,
\begin{equation}
\label{dnf}
\frac{d}{dt}(N-f(q))=\{H,N-f(q)\}=0.
\end{equation}
In quantum theory, it means that
\begin{equation}
\label{eigf}
[H,N-f(q)]=0.
\end{equation}
In other words, the Hamilton operator and the operator $(N-f(q))$ have a common set of eigenfunctions, and the basis in the Hilbert space is determined by a chosen gauge condition.

\section{Introduction of different reference frames in different spacetime regions}
Now we can consider a spacetime manifold that includes regions with different gauge conditions. First of all, let us note that the path integral approach enables one to study this situation. Imagine that the spacetime manifold consists of several regions ${\cal R}_1$, ${\cal R}_2$, ${\cal R}_3,$ \ldots, in
each of them various gauge conditions $C_1$, $C_2$, $C_3$, \ldots , being imposed. The regions are separated by boundaries ${\cal S}_1$, ${\cal S}_2$, \ldots\, For example, if ${\cal S}_1$ is the boundary between the regions ${\cal R}_1$ and ${\cal R}_2$, one has
\begin{eqnarray}
\label{PI1}
&&\hspace{-1cm}
\int\exp\left(iS\left[g_{\mu\nu}\right]\right)
 \prod\limits_{x\in{\cal M}}M\left[g_{\mu\nu}\right]
 \prod\limits_{\mu,\,\nu}dg_{\mu\nu}(x)\nonumber\\
&=&\int\exp\left(iS_{(eff)}\left[g_{\mu\nu},\,C_1,\,{\cal R}_1\right]\right)
 \prod\limits_{x\in{\cal R}_1}M\left[g_{\mu\nu},\,{\cal R}_1\right]
 \prod\limits_{\mu,\,\nu}dg_{\mu\nu}(x)\nonumber\\
&&\times\exp\left(iS_{(eff)}\left[g_{\mu\nu},\,C_2,\,{\cal R}_2\right]\right)
 \prod\limits_{x\in{\cal R}_2}M\left[g_{\mu\nu},\,{\cal R}_2\right]
 \prod\limits_{\mu,\,\nu}dg_{\mu\nu}(x)
 \prod\limits_{x\in{\cal S}_1}M\left[g_{\mu\nu},\,{\cal S}_1\right]
 \prod\limits_{\mu,\,\nu}dg_{\mu\nu}(x)\times\ldots
\end{eqnarray}
Where $S_{(eff)}[g_{\mu\nu},\ldots]$, $M[g_{\mu\nu},\ldots]$ are the effective action and the measure in the region indicated. The both depend on gauge conditions in the region.

From a theoretical point of view, the path integral (\ref{PI1}) enables one to consider spacetimes with non-trivial topology using various coordinates in different regions. However, there exist a problem if boundaries between regions are not a spacelike hypersurfaces of equal time. To derive the Schr\"odinger equation, we start from the relation between the wave function at some moment of time and the wave function at the previous moment. The assumption about an arbitrary topology prevents from introducing a global time in the whole spacetime, one should rather consider different clocks in every region.

In this section, we shall discuss a simplified situation depicted at Figure 1. The hypersurfaces
${\cal S}_0$, ${\cal S}_1$, \ldots\, correspond to some time moments $t_0$, $t_1$, \ldots. One can say that the topology is again assumed to be $\mathbb R\times\Sigma$, but we shall see that even in this case we shall come to results that could not be obtained in the Dirac -- Wheeler -- DeWitt approach.

\begin{figure}
\centering
\includegraphics[width=7 cm]{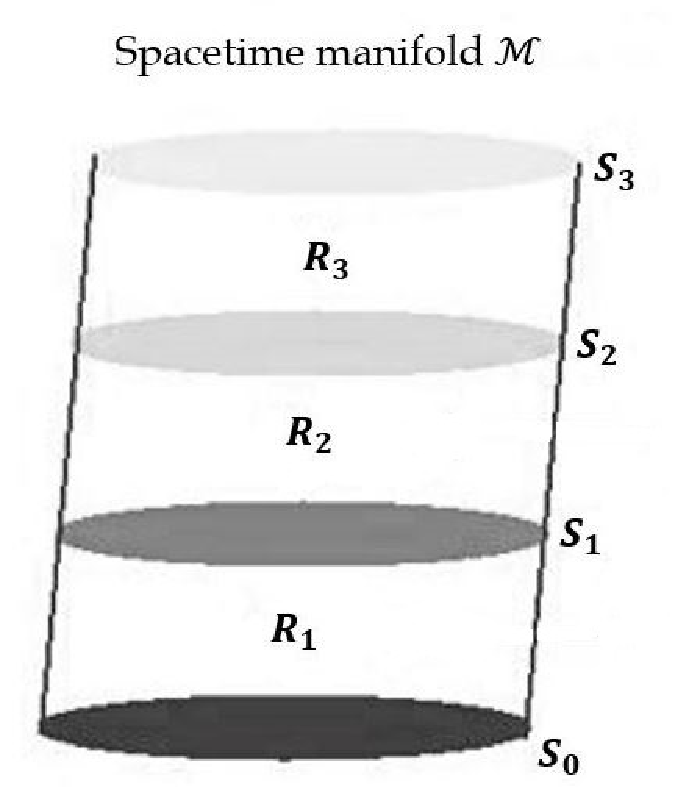}
\caption{\protect\small The schematic picture of a manifold that includes spacetime regions with different gauge conditions}
\label{fig1}
\end{figure}

Suppose that, in the region ${\cal R}_1$, the gauge conditions $C_1$ are imposed, and the physical Schr\"odinger equation with the Hamilton operator $H_{1(phys)}$ is valid; in the region ${\cal R}_2$, the gauge conditions $C_2$ are imposed, and the physical Schr\"odinger equation with the Hamilton operator $H_{2(phys)}$ is valid; etc. Denoting the quantum state on the surface ${\cal S}_0$ as
$|g_{\mu\nu}^{(0)},{\cal S}_0\rangle$, we obtain that the state on the surface ${\cal S}_1$ is
\begin{equation}
\label{S1}
|g_{\mu\nu}^{(1)},\,{\cal S}_1\rangle=
 \exp\left[-iH_{1(phys)}(t_1-t_0)\right]|g_{\mu\nu}^{(0)},\,{\cal S}_0\rangle.
\end{equation}

Within the region ${\cal R}_1$ the evolution of the physical system is governed by a unitary operator\linebreak
$\exp\left[-iH_{1(phys)}(t_1-t_0)\right]$. But, in the neighbour region ${\cal R}_2$, the Schr\"odinger equation with the Hamilton operator $H_{2(phys)}$ becomes active, and quantum states belong to another Hilbert space. One needs to expand the state (\ref{S1}) onto a basis constructed from eigenvectors of the operator $H_{2(phys)}$. Denote the operator of the transition to the new basis as
${\cal P}({\cal S}_1,\,t_1)$. Then, the initial state in the region ${\cal R}_2$ is
\begin{equation}
\label{projection}
{\cal P}({\cal S}_1,\,t_1)\exp\left[-iH_{1(phys)}(t_1-t_0)\right]|g_{\mu\nu}^{(0)},\,{\cal S}_0\rangle.
\end{equation}
Reasoning in this way, one would come to the conclusion that the evolution of the quantum state is described by a sequence of operators,
\begin{eqnarray}
\label{g3}
|g_{\mu\nu}^{(3)},{\cal S}_3\rangle
&=&\exp\left[-iH_{3(phys)}(t_3-t_2)\right]{\cal P}({\cal S}_2,\,t_2)\nonumber\\
&\times &\exp\left[-iH_{2(phys)}(t_2-t_1)\right]{\cal P}({\cal S}_1,\,t_1)
 \exp\left[-iH_{1(phys)}(t_1-t_0)\right]|g_{\mu\nu}^{(0)},\,{\cal S}_0\rangle.
\end{eqnarray}

In general, the operators ${\cal P}({\cal S}_i,\,t_i)$ are not unitary. They play the role of projection operators, which project states obtained by unitary evolution in a region ${\cal R}_i$ on a basis in Hilbert space in a neighbour region ${\cal R}_{i+1}$. So, at any border ${\cal S}_i$ between regions with different gauge conditions unitary evolution may be broken down.

Let us consider a simple example using the model with the effective action (\ref{S_fin}). Suppose that, in the region ${\cal R}_1$, the gauge condition $N=f(q)+k$ is imposed; its differential form is
$\dot N =\displaystyle\frac{\partial f}{\partial q^a}\dot q^a$, independently of the value of $k$. It means that we have chosen a basis of the Hilbert space that corresponds to this gauge condition. Then, the physical Hamilton operator is given by (\ref{Ham_phys}). The gauge condition in the neighbour region ${\cal R}_2$ is
$N=f(q)+\delta f(q)+k$, while $\delta f(q)$ being a small variation of the gauge fixing function $f(q)$. The physical Hamilton operator in the region ${\cal R}_2$ is
\begin{equation}
\label{Ham_phys_R2}
H_{(phys)}[f+\delta f]
 =\left[-\frac1{2M}\frac{\partial}{\partial q^a}\left(Mg^{ab}\frac{\partial}{\partial q^b}\right)
 +\left.U(N,q)+V[f+\delta f]\rule{0mm}{5mm}\right]\right|_{N=f(q)+\delta f+k}.
\end{equation}
Each of the operators $H_{(phys)}[f]$, $H_{(phys)}[f+\delta f]$ is Hermitian in a Hilbert space with a corresponding basis. On the other hand, taking into account that $\delta f$ is small, the operator (\ref{Ham_phys_R2}) can be presented as
\begin{equation}
\label{Ham_phys_approx}
H_{(phys)}[f+\delta f]=H_{(phys)}[f]+W[\delta f]+\delta U[\delta f]+V_1[\delta f],
\end{equation}
where
\begin{equation}
\label{W_op}
W[\delta f]
=\left.\left[\frac1{2M^2}\frac{\partial M}{\partial N}\delta f
  \frac{\partial}{\partial q^a}\left(MG^{ab}\frac{\partial}{\partial q^b}\right)
 -\frac1{2M}\frac{\partial}{\partial q^a}
  \left(\left(\frac{\partial M}{\partial N}g^{ab}+M\frac{\partial g^{ab}}{\partial N}\right)
  \delta f\frac{\partial}{\partial q^b}\right)\right]\right|_{N=f(q)+k}.
\end{equation}
One can check that $W[\delta f]$ is not Hermitian operator with respect to the basis in the region
${\cal R}_1$ with the gauge condition $N=f(q)+k$.

From this point of view, we can consider time-dependent gauge conditions. The path integral approach implies that one should approximate the effective action, including the gauge fixing term, at each small time interval $[t_i, t_{i+1}]$. We shall assume that, at each time interval, the alteration of the gauge condition $N=f(q)+k$ is
\begin{equation}
\label{alter}
\delta f_i(q)=\alpha f_i(q),
\end{equation}
where $\alpha$ is a small parameter. Then, the approximation of the gauge condition can be presented as a step function
\begin{equation}
\label{step}
N(t)=f(q)+\sum_{i=0}^n\alpha f_i(q)\theta(t-t_i)+k.
\end{equation}

It is worth noting that, at each time interval, the gauge condition (\ref{step}) does not depend on time. For example, at the interval $[t_n, t_{n+1}]$ one gets
\begin{equation}
\label{step_n}
N=f(q)+\sum_{i=0}^{n-1}\alpha f_i(q)+\delta f_n(q)+k.
\end{equation}
So, we have come to the case of a small variation of the gauge fixing function considered above. The small variation results in the appearance of a small correction to the Hamilton operator we had at the previous time interval. In the case of time-dependent gauge condition, it means that at every moment of time we have a Hamilton operator acting in its own ``instantaneous'' Hilbert space. The ``instantaneous'' Hamilton operator is a Hermitian operator at every moment of time, but it is non-Hermitian with respect to the Hilbert space that we had at the previous moment. There exist an analogy between the situation under consideration and particle creation in a non-stationary gravitational field, in the latter case, we also have an ``instantaneous'' Hamilton operator and an ``instantaneous'' Fock basis.

\section{Conclusions}
Let us now return to Eq.(\ref{g3}). It is worth comparing it with the formula describing the evolution of a quantum system according to von Neumann,
\begin{eqnarray}
\label{evol}
|\Psi(t_N)\rangle
&=& U(t_N,t_{N-1}){\cal P}(t_{N-1})U(t_{N-1},t_{N-2})\nonumber\\
&\times&\ldots U(t_3,t_2){\cal P}(t_2)U(t_2,t_1){\cal P}(t_1)U(t_1,t_0)|\Psi(t_0)\rangle.
\end{eqnarray}
As well known, von Neumann wrote \cite{Neumann} that there exist two ways of changing of quantum state of a physical system, namely, unitary evolution and changes as results of measurements under the physical system (reduction of the wave function). In (\ref{evol}), the projection operators ${\cal P}(t_i)$ correspond to measurement made at $t_1$, $t_2$, \ldots , $t_{N-1}$. The analogy between (\ref{g3}) and
(\ref{evol}) could be understood if we accept the interpretation of the reference frame as a measuring instrument representing the observer in quantum gravity. At the moments $t_1$, $t_2$, \ldots , $t_{N-1}$, transitions from one reference frame to another take place, and interaction between the measuring instrument (reference frame) and the physical subsystem changes. It makes us go to another Hilbert space.

However, we have already emphasize that, in general, the operators ${\cal P}(t_i)$ are not Hermitian. It leads to the following question:
\begin{itemize}
\item Can quantum gravity be the origin of non-unitarity?
\end{itemize}
This question remains open and require further investigations. Many physicists believe that unitarity is an inseparable property of any physical theory that cannot be broken down. On the other hand, in the framework of unitary evolution, it is not possible to describe irreversible processes we face all around. When one needs to describe such processes, one has to introduce some non-unitary operators artificially, so to speak, ``by hands''. In contrast, in the extended phase space approach to quantization of gravity, the appearance of the projection operators follows from the logical development of the accepted prerequisites.

It is important to remember that all the conclusions above are the consequences of the assumption about non-trivial topology and the absence of asymptotic states. These conclusions cannot be obtained in approaches based on the Wheeler -- DeWitt equation or making use of the assumption about asymptotic states.

\section*{Acknowledgments}
I am grateful to Prof. Sergey Sushkov and the Organizing Committee of RUSGRAV-18 for the invitation to give a plenary talk at the conference.

\end{document}